\documentclass{pasj00}
\draft

\def\ion#1#2{#1\,{\sc #2}}
\newcommand{\lam}{$\lambda$}

\input{epsf}

\usepackage{epsfig}

\begin{document}
\SetRunningHead{P.R. Young et al.}{Emission lines observed
  with Hinode/EIS} 

\title{EUV emission lines and diagnostics observed with Hinode/EIS}

\author{P. R. \textsc{Young},\altaffilmark{1}
        G. \textsc{Del Zanna},\altaffilmark{2}
        H. E. \textsc{Mason},\altaffilmark{3}
        K. P. \textsc{Dere},\altaffilmark{4}
        E. \textsc{Landi},\altaffilmark{5}
        M. \textsc{Landini},\altaffilmark{6}
        G. A. \textsc{Doschek},\altaffilmark{7}
        C. M. \textsc{Brown},\altaffilmark{7}
        J. L. \textsc{Culhane},\altaffilmark{2}
        L. K. \textsc{Harra},\altaffilmark{2}
        T. \textsc{Watanabe},\altaffilmark{8}
        H. \sc{Hara}\altaffilmark{8}}
\altaffiltext{1}{STFC Rutherford Appleton Laboratory, Chilton, Didcot,
  Oxfordshire, OX11 0QX, U.K.}
\altaffiltext{2}{University College London, Department of Space and Climate
  Physics, Holmbury St. Mary,
  Dorking, Surrey, UK}
\altaffiltext{3}{DAMTP, Centre for Mathematical Sciences, Cambridge,
  UK}
\altaffiltext{4}{George Mason University, 4400 University Dr., Fairfax
VA, 22030, USA}
\altaffiltext{5}{Artep, Inc at Naval Research Laboratory, 4555
        Overlook Ave. S.W., 20375-5320, Washington DC, USA}
\altaffiltext{6}{Dipartimento di Astronomia e Scienza dello Spazio, Universit\'a di Firenze, Largo E. Fermi 2, 50125 Florence, Italy}
\altaffiltext{7}{Code 7670, Naval
  Research Laboratory, Washington, DC 20375-5352,
  USA}
\altaffiltext{8}{National Astronomical Observatory of Japan, National
        Institutes of Natural Sciences, Mitaka, Tokyo 181-8588}


%


\maketitle

\begin{abstract}
Quiet Sun and active region spectra from the Hinode/EIS
instrument are presented, and the strongest lines from different
temperature regions discussed. A list of emission lines
recommended to be included in EIS observation studies is presented
based on analysis of blending and diagnostic potential using the
CHIANTI atomic database. In addition we identify the most useful
density diagnostics from the ions covered by EIS.
\end{abstract}

\section{Introduction}

The EUV Imaging Spectrometer (EIS) on board Hinode \citep{kosugi07}
takes high 
resolution spectra in the two wavelength bands 170--211~\AA\ and
246--292~\AA, referred to here as the short and long wavelength bands
(SW and LW, respectively). The instrument is described in detail by
\citet{culhane07}. The two wavelength bands were chosen as
they contain excellent diagnostics of coronal and flaring plasma; the
SW band in particular is the most rich 
in coronal plasma diagnostics in the whole EUV region, and EIS is the
first satellite-based instrument to observe it in high resolution.

Constraints due to on board storage and telemetry mean that
for EIS observations with a reasonable cadence and spatial coverage it
is not possible to send the complete spectra to Earth. An on board selection of particular emission
lines is thus performed, the reduced telemetry consisting of a set of
windows $w$ pixels wide by 
$h$ pixels high centred on each of the selected wavelengths. 
The most important aspect of designing an EIS
observation study is to choose the emission lines for the
science you want to do.
The present paper provides a guide to the most prominent
and useful emission lines in the EIS spectra based on early data analysis, and also highlights key
density diagnostics.
Pre-flight descriptions of the EIS capabilities, including diagnostics
and line identifications, are given in 
\citet{delzanna_mason05_eis}.

\section{Instrument capability}

\begin{figure*}[h]
\centerline{\epsfxsize=17.5cm\epsfbox{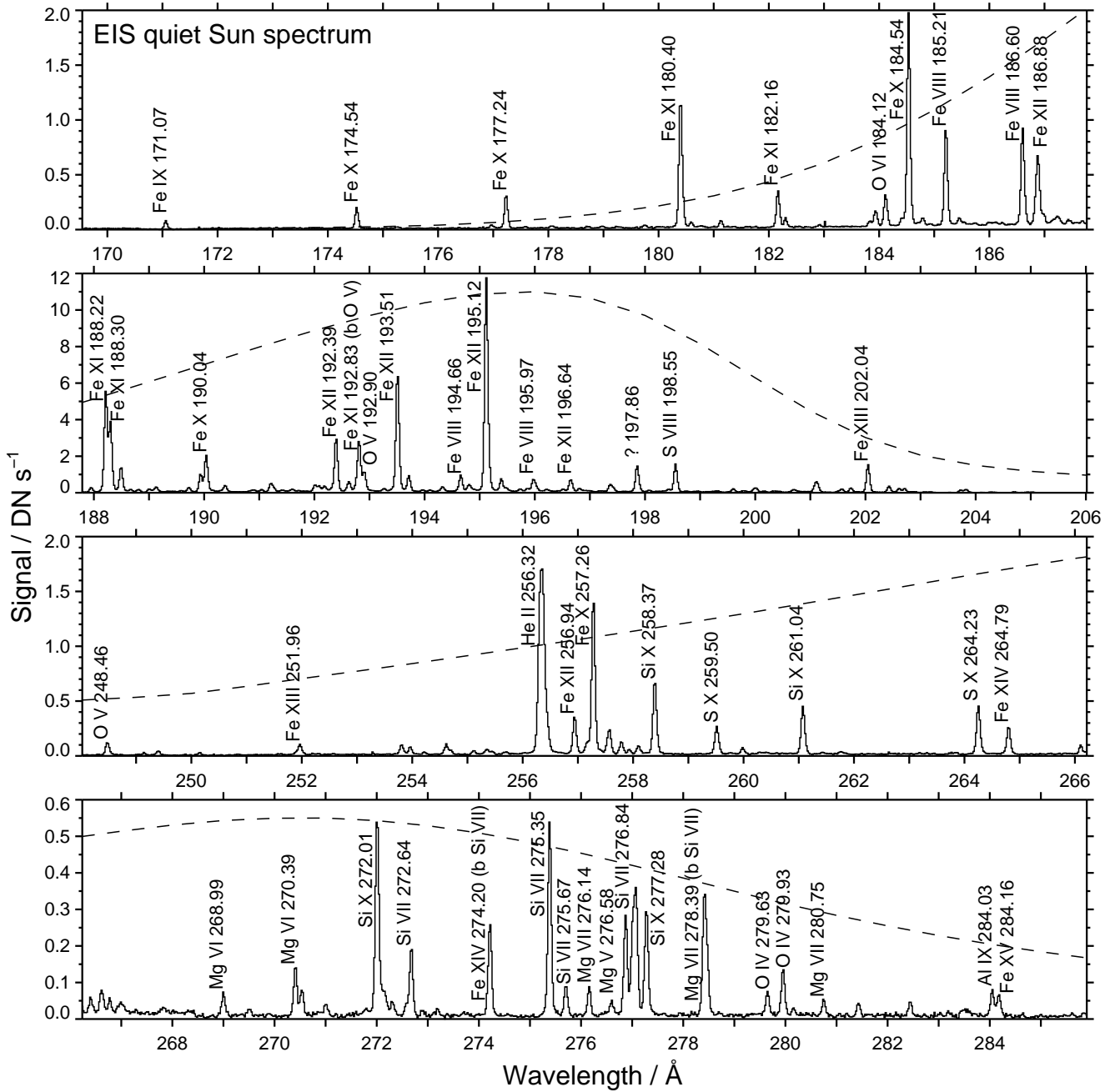}}
\caption{An EIS quiet Sun spectrum obtained on 2006 December 23 with
identifications of important lines shown. The dashed line shows the
effective area of the instrument, which has a peak value of
0.31~cm$^2$ in the SW band and 0.11~cm$^2$ in the LW band.}
\label{fig.quiet}
\end{figure*}

\begin{figure*}[h]
\centerline{\epsfxsize=17.5cm\epsfbox{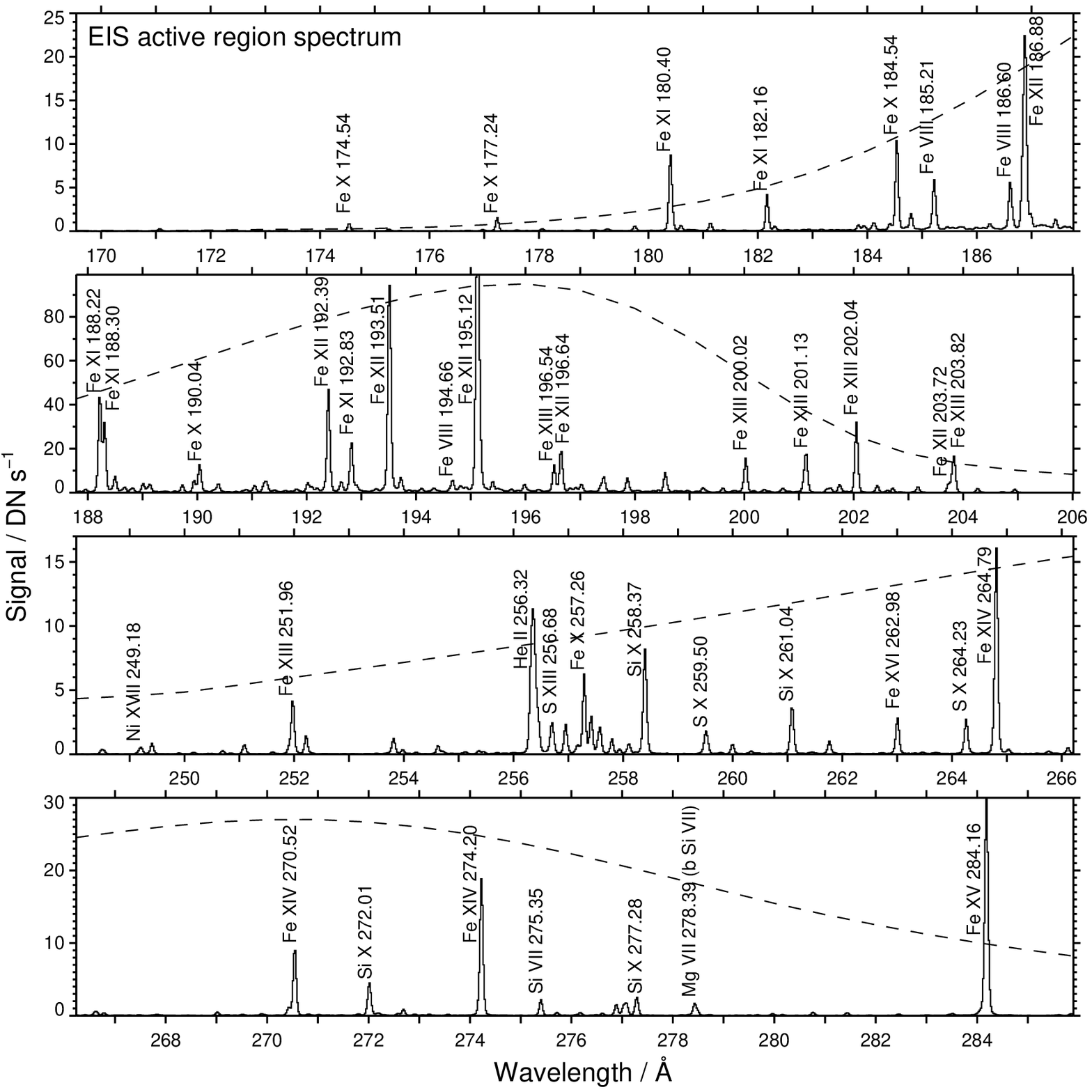}}
\caption{An EIS active region spectrum obtained on 2006 November 4 with
identifications of important lines shown. The dashed line shows the
effective area of the instrument, which has a peak value of
0.31~cm$^2$ in the SW band and 0.11~cm$^2$ in the LW band.}
\label{fig.active}
\end{figure*}

The EIS instrument is performing very well in terms of expected
sensitivity and spectral resolution, and the observed spectra are very
similar to pre-flight predictions made by \citet{delzanna_mason05_eis}
using the CHIANTI database \citep{landi_etal:06,dere97} and the expected
instrument parameters. 
Figs.~\ref{fig.quiet} and \ref{fig.active} show sample averaged
spectra obtained with EIS for quiet and active regions observed on
2006 December 23 16:10~UT and 2006 November 4 11:49~UT, respectively.
The 1${^\prime}{^\prime}$ slit was used in each case. The full width
at half maximum of the emission lines is around 0.065--0.075~\AA\
(3--3.5 pixels), corresponding to a spectral resolution of
$\approx$~3000--4,000.

A key factor when judging the usefulness of an EIS emission line is the
telescope effective area (EA) at that wavelength, which determines the
fraction of incident photons that arrive at the detector. The EA
curves are overplotted as dashed lines on the spectra in
Figs.~\ref{fig.quiet} and \ref{fig.active}. The use of two different multilayer coatings
on the EIS optical surfaces leads to EA curves that are peaked for the
two channels, with less sensitivity at the ends of each wavelength range.
This is particularly so for the SW channel which peaks at 
196~\AA,
very close to the strong \lam195.12 line of \ion{Fe}{xii}, making this
the strongest line observed by EIS in most conditions. The
\ion{Fe}{ix} \lam171.02 and \ion{Fe}{x} \lam174.53 lines are
comparable in strength to \ion{Fe}{xii} \lam195.12 in moderately
active solar conditions \citep{malinovsky73}, but when observed with
EIS they are factors 1000 and 200 times weaker than \lam195.12,
respectively. The EA curve for the long wavelength (LW) band is less
peaked giving a more consistent instrument sensitivity across the
band. The shape of these EA curves is an important consideration when
choosing emission lines for EIS observation studies: a weak line near
195~\AA\ can yield more instrument counts than a stronger line at the
edges of the EA curve. Specific examples of this are discussed below.

Another important factor when considering emission lines is the degree
of blending. Even with the high spectral resolution of EIS, many lines
are blended with other species, particularly in the SW channel. 
Examples of such blends are described below and 
include the blending of two of the three EIS `core lines' --
\ion{He}{ii} \lam256.32  
and \ion{Ca}{xvii} \lam192.82 -- which are lines included by default
in all EIS observation studies. In many cases the contributions of
blending lines can be estimated quite accurately by making use of line
ratios that are insensitive to the plasma conditions and again
examples are given below.

The ions discussed in the following sections we believe will be sufficient
for most science studies. However, there are a number of interesting
lines from ions of nickel, sulphur, argon and calcium that will be
valuable for abundance or flare studies. Discussion of these lines
will be deferred to a future paper.

All atomic data used in compiling this paper are from v.5.2 of
the CHIANTI atomic database \citep{landi_etal:06, dere97} and, in
particular, all wavelengths given below are from
CHIANTI. 
When referring to
intensity ratios below, ratios in energy units (rather than photon
units) are implied. 
When referring to the strength of EIS lines, we usually use the measure
``data number'' (DN), which is the unit for the data stored in the
level 0 EIS FITS files. The number of photons is obtained by
multiplying the DN by a wavelength varying factor of around 2--3.
Due to space restrictions, we can not give the transition
identifications for all the lines in the current document. For these
we refer the reader to \citet{delzanna_mason05_eis} or the CHIANTI
database. 

\section{Iron lines}

\subsection{Fe VIII}

All \ion{Fe}{viii} lines are found in the SW band, and the
strongest lines are at 185.21, 186.60 and 194.66~\AA\ lines, which are
all comparable in strength in terms of DN. \lam185.21 is blended with
the \ion{Ni}{xvi} 185.23~\AA\ ($\log\,T=6.4$) line which becomes apparent 
in active region intensity maps with a ``mist'' of emission visible
around the hot core of the active region. 
The \ion{Ni}{xvi} contribution can be estimated by 
observing the  \ion{Ni}{xvi} 195.27~\AA\ line (which lies very close
to \ion{Fe}{xii} \lam195.12) as the \lam195.27/\lam185.23 intensity
has a fixed value of 0.23.
\lam186.60 is blended with \ion{Ca}{xiv} \lam186.61 ($\log\,T=6.5$),
and the \ion{Ca}{xiv} contribution can be estimated by measuring the
\lam193.86 line: the \lam186.61/\lam193.86 intensity ratio being fixed
at 0.69.
Note that \ion{Fe}{viii} is typically found
to be strongest in the footpoints of loops at the edges of active
regions, where hot ions such as \ion{Ni}{xvi} and \ion{Ca}{xiv} are
often negligible.  

\ion{Fe}{viii} \lam194.66 is not directly blended but has an unidentified line
in the long wavelength wing at $\approx$~194.80~\AA. \lam194.66 may
possibly be affected if the \ion{Fe}{xii} \lam195.12 line shows a high
velocity blue-shifted component, or is broadened significantly. In
summary, we thus recommend observing both the \lam185.21 and
\lam194.66 lines -- the former is in a relatively clean part of the
spectrum (apart from the high temperature \ion{Ni}{xvi} line) making
it good for velocity studies, and the latter can be used in active
regions when \ion{Ni}{xvi} significantly contaminates the \lam185.21 line.
The \ion{Fe}{viii} lines show density sensitivity below
$10^8$~cm$^{-3}$ and so are not generally useful; the
\lam186.60/\lam185.21 ratio is the best diagnostic in these conditions,
however. 


An important point to note about \ion{Fe}{viii} is that it seems to be
formed at $\log\,T=5.8$ rather than the temperature of $\log\,T=5.6$ predicted
by \citet{mazzotta98}. This is discussed further in \citet{young07b}.

\subsection{Fe\,X }

The strongest \ion{Fe}{x} lines observed by EIS 
are at 184.54, 190.04 and 257.26~\AA\ (the
latter actually a self-blend of two lines). 
\lam184.54 and \lam190.04  are
comparable in strength in terms of DN. \lam190.04  is partly
blended with an unknown line at around 189.94~\AA\
which seems to be formed around $\log\,T = 5.8$--5.9.
The \lam190.04/\lam184.54 ratio is insensitive to density, but \lam257.26
is sensitive to density when taken relative to either \lam190.04 or
\lam184.54. The \lam257.26/\lam184.54 ratio is recommended and shown in
Fig.~\ref{fig.dens}. \lam257.26 is in a crowded part
of the spectrum 
and a large wavelength window (at least 50 pixels) is required to
obtain a good estimate of the spectrum background.

\subsection{Fe\,XI}\label{sect.fe11}

\ion{Fe}{xi} is a very complex ion and even line identifications of
strong lines are uncertain. Improvements in computing power and
techniques should yield accurate atomic data in the near
future, and work is underway by P.J.~Storey in collaboration with some
of the present authors.

The strongest \ion{Fe}{xi} lines observed by EIS are partly blended
with each other and found at 188.23 and 188.30~\AA. Although the
identification of 
\lam188.30 is uncertain, the \lam188.23/\lam188.30 ratio is found to
be nearly constant in the EIS data with a ratio of around 0.7, suggesting
they are from the same ion. We believe the 3s$^2$3p$^4$ $^3$P$_{2,1}$
-- 3s$^2$3p$^3$($^2$D)3d $^3P^{\rm o}_{2}$ transitions give rise to
\lam188.23 and 
another line at 192.83~\AA, respectively.  This latter line provides a
component to the complex blend at this wavelength which is discussed
further in Sects.~\ref{sect.o5} and \ref{sect.ca17}.

The \ion{Fe}{xi} lines at 180.40 and 182.16~\AA\ are well identified
and density sensitive relative to each other, but the EA
is low in this region. The \lam182.16/\lam188.23 density sensitive
ratio is preferred, 
with the proviso that the \lam188.23 transition identification is
uncertain.

\subsection{Fe\,XII }

Many discrepancies between theory and observations for \ion{Fe}{xii}
have been resolved recently following a new atomic calculation 
\citep{del_zanna_mason05_fexii, storey_etal:05}, and we can now
have confidence in 
using the ion for diagnostic work.

The three strongest \ion{Fe}{xii} lines are the decays of the 
3s$^2$3p$^2$($^3$P)3d $^4$P$_{5/2,3/2,1/2}$ states to the ground state,
giving lines at 195.12, 193.51, 192.39~\AA, respectively.
\lam195.12 lies at the peak of the EIS
sensitivity curve and so
is the strongest emission line observed by EIS in most conditions. It
is one of the EIS core lines and therefore included in all EIS studies.
\lam195.12 has been found to be broader than the \lam193.51 line which
could be  indirect confirmation of the identification proposed
in \citet{del_zanna_mason05_fexii} of a \ion{Fe}{xii} line at
195.18~\AA. At $10^{10}$~cm$^{-3}$ this line is predicted to
contribute 10~\% to the feature at 195.1~\AA.

The only problem with observing the \lam195.12 line is that it is likely
to saturate on the detector if long exposure times are used. In active
conditions even a 30 second exposure can lead to saturation 
in bright parts of an active
region. We thus recommend that the \lam192.39 or \lam193.51 lines are
observed in addition to \lam195 as they are around 27~\% and 60~\%
weaker in terms of DN. 

\ion{Fe}{xii} provides some of the best density diagnostics 
for EIS. The ratio of either of the  \lam196.64 or \lam186.88 lines to any 
of \lam\lam195.12, 193.51, 192.39 
is sensitive to a wide range of densities
($\log\,N_{\rm e} = 8$--12). \lam186.88 
is a blend of two \ion{Fe}{xii} lines,
and is stronger than \lam196.64 in terms of DN,
but it is also blended with a \ion{S}{xi} transition at 186.84~\AA. This
generally makes a small contribution to the \ion{Fe}{xii} line, and it can
be accurately assessed by measuring \ion{S}{xi} \lam191.27 as
\lam186.84/\lam191.27 has a fixed ratio of 0.20. \lam196.64 lies
close to \ion{Fe}{xiii} \lam196.54 (another recommended line -- see
below) but can generally be resolved, and thus we recommend this line
for the \ion{Fe}{xii} density diagnostic.

\subsection{Fe\,XIII }

The three most important \ion{Fe}{xiii} lines are at 196.54, 202.04
and 203.82~\AA. They form the best coronal density
diagnostics available to EIS due to their high sensitivity to density
(Fig.~\ref{fig.dens}). \lam203.82 is  a blend 
of two \ion{Fe}{xiii} lines at 203.80 and 203.83~\AA\ that are
approximately in the ratio 1:3. Interpretation of \lam203.82
is hampered by a blend with \ion{Fe}{xii} \lam203.72, but fitting the
combined feature with two Gaussians can usually separate the
\ion{Fe}{xii} and \ion{Fe}{xiii} components. The \lam202.04 line is
unblended, while \lam196.54 is easily resolvable from \ion{Fe}{xii}
\lam196.64 in most conditions. The \lam196.54/\lam202.04 ratio is more
sensitive to 
density than \lam203.82/\lam202.04 above around
$10^{10}$~cm$^{-3}$ and is also sensitive to higher densities, but
both are highly recommended to be included in EIS studies.

Other  strong \ion{Fe}{xiii} lines are found at 197.43~\AA,
201.13~\AA\ (blended with \ion{Fe}{xii} \lam201.14)
and 251.96~\AA, but these are not as useful as the aforementioned lines.

\subsection{Fe\,XIV }

A number of prominent \ion{Fe}{xiv} lines are found in the EIS
wavebands, and the one recommended here is at 274.20~\AA. Although
there is a blend with \ion{Si}{vii} \lam274.18, this can be quantified
if \ion{Si}{vii} \lam275.35 (one of the recommended lines) is
also observed since the 
\lam274.18/\lam275.35 ratio is at most 0.25. In most active region
conditions the blend can safely be ignored.  The \ion{Fe}{xiv}
\lam264.79 line
 yields a good density diagnostic 
 relative to \lam274.20 (Fig.~\ref{fig.dens}) and is
recommended for probing hotter parts of active regions.
Another strong line is \lam270.52, but this is weaker than \lam274.20 in
all conditions, and the \lam264.79/\lam270.52 ratio is less sensitive
to density.

\subsection{Fe\,XV }

The \ion{Fe}{xv} 284.16~\AA\ line dominates this region of the EIS spectrum in
active conditions and is the
strongest line from the ion. We recommend its inclusion in any EIS study.
Note that in quiet Sun conditions the line is very weak or
non-existent and an \ion{Al}{ix} line at 284.03~\AA\ becomes apparent.

\subsection{Fe\,XVI } 

Three lines are found in the EIS long wavelength band at 251.06, 262.98
and 265.00~\AA. They are all temperature and density insensitive
relative to each other, \lam262.98 
being the strongest. This
latter line is unblended and  is recommended for
inclusion in all observations.

\subsection{Fe\,XVII}

\ion{Fe}{xvii} is formed over a broad range of temperatures, with its
maximum abundance at $\log\,T=6.6$. The strongest \ion{Fe}{xvii}
transitions are found at X-ray 
wavelengths but there are a number of weak transitions in the EIS
bands, particularly the LW band. The strongest of the lines is at
254.87~\AA\ and, although it is much weaker than \ion{Ca}{xvii}
\lam192.82 (which is formed at a similar temperature), it does not
suffer the blending problems of this line. Note that v.5.2 of
CHIANTI does not give the correct wavelength for this transition,
instead listing it at 254.35~\AA. The \ion{Fe}{xvii} model is being
reassessed by the CHIANTI team, and caution should be applied if using
this line for quantitative analysis.

\subsection{Fe\,XXIII}

The 
\lam263.76 line is predicted to be the strongest line during large flares,
after the \ion{Fe}{xxiv} lines, and should be unblended. All EIS
studies for active regions, 
microflares and flares are recommended to include this line.

\subsection{Fe\,XXIV}

The Li-like doublet 
2s $^2$S$_{1/2}$ -- 2p $^2P^{\rm o}_{3/2,1/2}$ 
gives rise to lines at 192.03 and 255.11~\AA, respectively,
and during large flares they  become the 
most prominent lines in the EIS spectra.
\lam192.03 is blended with another line, believed to be from
\ion{Fe}{xi}, while
\lam255.11 is  blended with a 
relatively weak \ion{S}{x} line at 255.06~\AA\ ($\log\,T=6.1$), thus
care must be taken when searching for a weak signal in the
\ion{Fe}{xxiv} lines. Although the intensity of \lam192.03 is
predicted to be 2.5 times stronger than \lam255.11, it will be around
12 times stronger in terms of DN due to the larger EA at 192~\AA.
Note that the
\ion{Fe}{xxiv} lines have been seen in observations of C class flares by
EIS, and are recommended for all active region studies.


\section{Oxygen lines}

\ion{O}{iv--vi} each have very strong 2s--2p transitions at longer UV
wavelengths, beyond the reach of EIS. However, these ions have  a number of
weak $n=2$ to $n=3$ transitions in the EIS wavebands that are valuable
diagnostics of the transition region.

\subsection{O\,IV}

There are many weak lines from \ion{O}{iv} predicted in the SW and LW
bands. The strongest is found at 279.93~\AA, and there is a nearby
line at 279.63~\AA\ that is a factor 2 weaker. These \ion{O}{iv} lines
are weaker than the lines of \ion{O}{v} and \ion{O}{vi} mentioned
below, and so the latter are generally preferred unless the transition
region is specifically the region of interest.

\subsection{O\,V}\label{sect.o5}

The strongest \ion{O}{v} lines arise from the 2p
$^3P^{\rm o}_J$ -- 3d $^3$D$_{J^\prime}$ multiplet and are found between
192.75 and 192.91~\AA, contributing to one of the most difficult parts
of the EIS spectrum to analyse. The EIS core line \ion{Ca}{xvii}
\lam192.82 ($\log\,T=6.7$) is in this region, as well as \ion{Fe}{xi}
\lam192.83. 
There are six \ion{O}{v} transitions in all from the multiplet, although
one has negligible intensity. The remaining five lines give rise to
three features at 192.75~\AA, 192.80~\AA, and 192.90~\AA\ (the latter
two comprising of two co-incident transitions each). There is density
sensitivity amongst the lines, but they lie in the approximate ratio
0.16\,:\,0.4--0.45\,:\,1.0 in most conditions -- see \citet{young07b}. 
\lam192.80  is blended with the \ion{Fe}{xi} and \ion{Ca}{xvii}
lines, with \lam192.75 discernible in the short wavelength wing. The
\lam192.90 component can usually be resolved and thus by measuring
this line one can estimate the \ion{O}{v} contribution to the
\ion{Fe}{xi}--\ion{Ca}{xvii} feature.

The other useful \ion{O}{v} line is at 248.46~\AA\ which is  around ten
times weaker than the \lam192.90 line due to the much lower EA at this
wavelength. The \lam192.90/\lam248.46 ratio is sensitive to densities
above $10^{10}$~cm$^{-3}$ (Fig.~\ref{fig.dens}) and thus will be useful for the study of the
active region transition region brightenings discussed by \citet{young07b}.
CHIANTI lists a blend with
\ion{Al}{viii} \lam248.46 ($\log\,T=6.0$) which should only be a minor
component in most conditions. However, its contribution can be
estimated from the \ion{Al}{viii} \lam248.46/\lam250.14 intensity
ratio which has a fixed value of 0.60.

\subsection{O\,VI}

Two \ion{O}{vi} lines are available at 183.94 and 184.12~\AA\
and the ratio \lam183.94/\lam184.12 is 0.5. They lie close to the
recommended line \ion{Fe}{x} \lam184.54 and a single broad wavelength
window can be used to pick up all three lines. 

\section{Magnesium lines}

\subsection{Mg\,V}

Only one line is found in the EIS wavelength bands at
276.58~\AA. While very weak in most circumstances, the line is
significantly enhanced in loop footpoints \citep{young07b} and gives
valuable temperature information.

\subsection{Mg VI}

Two lines are found at 268.99 and 270.39~\AA\ which are in the approximate
ratio of 1:2 in most conditions.
\lam270.39 lies in the wing of \ion{Fe}{xiv} \lam270.52 which is usually much
stronger in active region conditions. \lam268.99  is
unblended. Both lines can be strongly enhanced in loop footpoints as
demonstrated in \citet{young07b}. Note that
\lam268.99  is relatively isolated and thus valuable for identifying
loop footpoints in 40$^{\prime\prime}$ slot data.

\subsection{Mg\,VII }

The 2s$^2$2p$^2$ $^3$P$_J$ -- 2s2p$^3$ $^3S^{\rm o}_1$ multiplet is found at
276.14, 276.99 and 278.39~\AA, with the lines lying in the ratio
0.20\,:\,0.60\,:\,1.0. The strongest line, \lam278.39, is blended
with \ion{Si}{vii} \lam278.44 but they can be resolved either through
a double Gaussian fit, or by using the \ion{Si}{vii}
\lam278.44/\lam275.35 insensitive ratio \citep{young07b}.
\lam276.99 is blended with \ion{Si}{viii} \lam277.04, but
\lam276.14 is unblended.
A further \ion{Mg}{vii} line is found at 280.75~\AA\ which is
unblended and of great value as a density diagnostic relative to any
of the other three lines, being sensitive in the range
$10^8$--$10^{11}$~cm$^{-3}$. The ratio is used
by \citet{young07b} to 
measure the density in loop footpoints.

\section{Silicon lines}

\subsection{Si\,VII}

The six 2s$^2$2p$^4$ $^3$P$_J$ - 2s2p$^5$ $^3P^{\rm o}_{J^\prime}$
transitions are found in the 
EIS long 
wavelength band, and the strongest is at 275.35~\AA\  which is
unblended. Note that, from visual inspection of EIS images,
\ion{Si}{vii} is formed at 
around the same temperature as \ion{Fe}{viii} but is weaker by a
factor two in terms of DN
compared to the strongest lines from that ion. 

\subsection{Si\,X}\label{sect.si10}

The six lines belonging to the 2s$^2$2p $^2P^{\rm o}_J$ -- 2s2p$^2$
$^2$P$_{J^\prime}$ 
and 2s$^2$2p $^2P^{\rm o}_{J}$ -- 
2s2p$^2$ $^2$S$_{1/2}$ transitions are found in the LW channel, with the
strongest being \lam258.37 and \lam261.04 which form a density
diagnostic (Fig.~\ref{fig.dens}). Both lines appear to be unblended.
\ion{Si}{x} \lam256.37 is blended with \ion{He}{ii}
\lam256.32, hampering interpretation of this key line, however
the \lam256.37/\lam261.04 ratio has a fixed intensity ratio of 0.89,
allowing the \ion{Si}{x} contribution to be estimated.

According to the \citet{mazzotta98} ion balance calculations,
\ion{Si}{x} is formed at almost exactly the same temperature as
\ion{Fe}{xii}, and so the \lam258.37/\lam261.04 density diagnostic
should probe the same region 
as \ion{Fe}{xii} \lam196.64/\lam195.12. Although \ion{Fe}{xii} \lam196.64 has 
approximately the same DN as  \ion{Si}{x} \lam258.37,
\ion{Fe}{xii} \lam195.12 is much stronger than \ion{Si}{x} \lam261.04
and so the \ion{Fe}{xii}  
ratio is to be preferred.

\section{Other ions}

\subsection{He\,II}

\ion{He}{ii} \lam256.32 is the coolest line observed by EIS and
also the strongest line formed below $10^6$~K, so it was selected
as one of the three EIS core lines to appear in every EIS study. It is
actually a self blend of two transitions with almost identical
wavelengths. 
Interpretation of \lam256.32  is complicated by blends with
\ion{Si}{x} \lam256.37 (see 
Sect.~\ref{sect.si10}), \ion{Fe}{xiii} \lam256.42 and \ion{Fe}{xii}
\lam256.41. For disk observations of the quiet Sun and active regions,
\ion{He}{ii} should dominate the feature contributing 80~\% or more to
the blend, but above the limb the coronal lines will generally
dominate. If the \ion{He}{ii} line is crucial to your science then we
recommend taking other \ion{Si}{x}, \ion{Fe}{xii} and \ion{Fe}{xiii}
lines in order to correctly estimate their contributions. Despite
these blends, \lam256.32 is still very useful for the study of
explosive events and other dynamic phenomena.

\subsection{S\,XIII}

The strongest line of the \ion{S}{xiii} spectrum is found at
256.68~\AA\ and is comparable in strength to \ion{Fe}{xvi} \lam262.98
which is formed at the same temperature. There is a contribution from
\ion{Ni}{xvi} \lam256.62, but this can be estimated from the
\lam256.62/\lam195.28 intensity ratio 
which is approximately 0.28 in all conditions.  
The \ion{S}{xiii} line is also close to \ion{He}{ii} \lam256.32
and so can be affected if the latter is broadened or redshifted
in dynamic events.

\subsection{Ni\,XVII}

The 
\lam249.18 line of \ion{Ni}{xvii} is the analogous transition to
\lam284.16 of 
the iso-electronic \ion{Fe}{xv} ion.
The line is
unblended and a valuable probe of the hot cores of active regions.
It is weaker than  \ion{Fe}{xvi} \lam262.98 and 
\ion{S}{xiii} \lam256.68, but according to the 
\citet{mazzotta98} ion balance calculations this line should
be formed in slightly hotter plasma, in the $\log\,T=6.4-6.5$ range.

\subsection{Ca\,XVII}\label{sect.ca17}

\ion{Ca}{xvii} is formed at $\log\,T=6.7$ and its strongest line
is found in the EIS SW band at 192.82~\AA, where the instrument is
very sensitive. This line was selected as one of the three EIS core
lines to ensure at least one flare line is present in all EIS studies.
However, \lam192.82 is
part of a complex blend comprising five
lines of \ion{O}{v} and two lines of \ion{Fe}{xi} and so requires
careful analysis. The \ion{O}{v} lines were discussed in
Sect.~\ref{sect.o5} where a method of estimating the contribution from
\ion{O}{v} was given.

Using CHIANTI to estimate the contribution of \ion{Fe}{xi} is somewhat
uncertain due to the known problems with line identifications and
atomic data for this ion. There are two lines given by CHIANTI 
at wavelengths 192.830 and  192.832~\AA. The latter is a theoretical
wavelength and thus uncertain. \lam192.830 has a fixed ratio relative
to the strong \lam188.23 line (Sect.~\ref{sect.fe11}), with a
theoretical intensity ratio of 
0.21. \lam192.832 is predicted to be weaker than \lam192.830, the
ratio \lam192.832/\lam192.830 varying from 0.20 at low density to 0.55
at high density.
Inspecting  EIS spectra taken
above the solar limb in quiet Sun conditions (where both \ion{O}{v}
and \ion{Ca}{xvii} are negligible), we find the observed
\lam192.83/\lam188.23 ratio to be around 0.26, which is consistent
with both the CHIANTI \lam192.830 and \lam192.832 lines contributing
to the observed feature at 192.83~\AA. Our recommended prescription
for estimating the \ion{Fe}{xi} contribution to the \ion{Ca}{xvii}
line is thus to measure the \ion{Fe}{xi} \lam188.23 line and multiply
the intensity by the factor 0.26 to obtain the \lam192.83 intensity.

Despite the difficulties with blending it should be noted that
\ion{Ca}{xvii} completely dominates the 
other lines in large flares and thus is an important line for such
observations.

\begin{figure*}[h]
\centerline{\epsfxsize=15cm\epsfbox{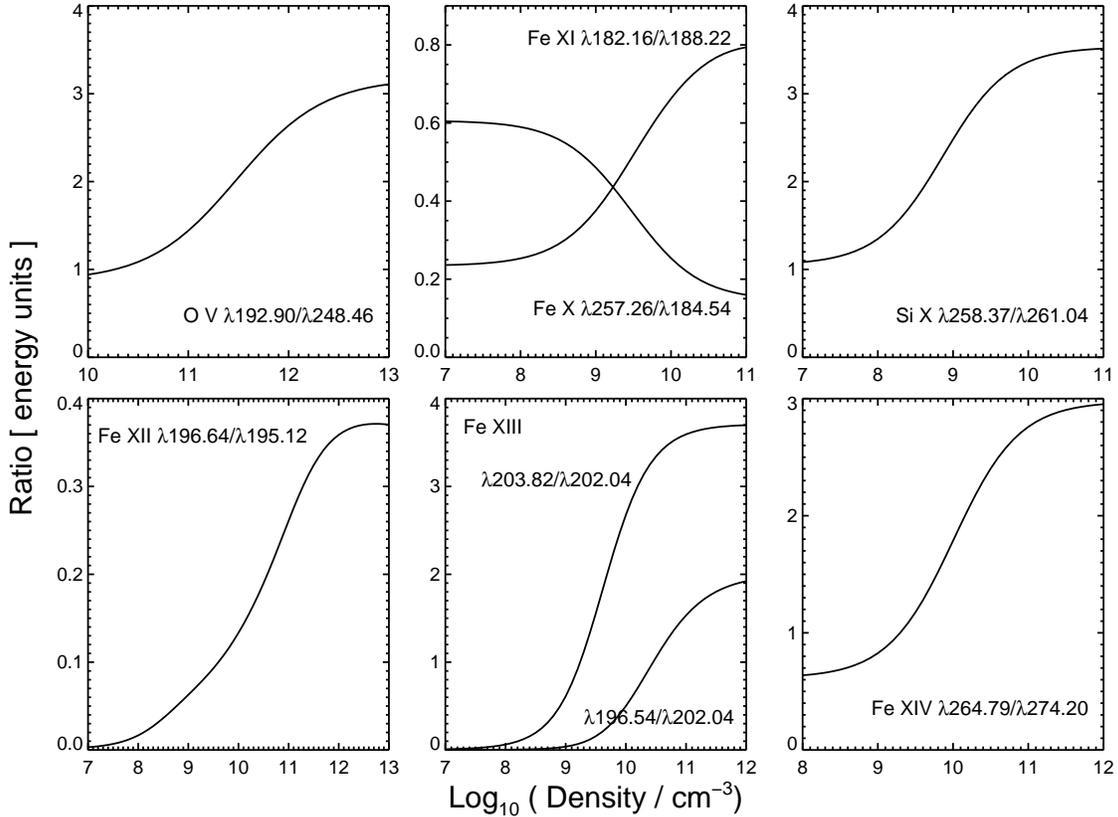}}
\caption{Recommended density diagnostic emission line ratios
  discussed in the text. Plots derived using atomic data and
  software from v.5.2 of the CHIANTI atomic database. Note that the
  other recommended diagnostic -- 
  \ion{Mg}{vii} \lam280.78/\lam278.39 --  is presented in
  \citet{young07b}.} 
\label{fig.dens}
\end{figure*}

\begin{table}[h]
\caption{EIS emission lines recommended to be included 
in observation studies. A `b' indicates blending (see text).}
\begin{center}
\begin{tabular}{llll}
\hline
\hline
\noalign{\smallskip}
&Ion & Wavelength / \AA &$\log\,T_{\rm max}$ \\
\hline
\noalign{\smallskip}
\multicolumn{4}{l}{Cool lines ($\log\,T_{\rm max} < 6.0$)} \\
\noalign{\smallskip}
& \ion{He}{ii} & 256.32 b & 4.7 \\
& \ion{O}{v}  & 192.90 & 5.4 \\
&             & 248.46 b & 5.4 \\
& \ion{O}{vi} & 184.12 & 5.5 \\
& \ion{Mg}{v} & 276.58 &5.5 \\
& \ion{Mg}{vi} &268.99 & 5.7 \\
& \ion{Mg}{vii} &278.39 b & 5.8 \\
&               &280.75 &5.8 \\
& \ion{Si}{vii} &275.35 & 5.8 \\
& \ion{Fe}{viii} &185.21 & 5.8 \\
&  &194.66 & 5.8 \\
\noalign{\smallskip}
\multicolumn{4}{l}{Coronal lines ($6.0\le \log\,T_{\rm max} <
  6.4$)}\\
\noalign{\smallskip}
& \ion{Fe}{x} & 184.54 & 6.0 \\
&             & 257.26 & 6.0 \\
& \ion{Fe}{xi} & 188.23 & 6.1 \\
&              &182.16 & 6.1 \\
& \ion{Si}{x} & 258.37 & 6.1 \\
&             & 261.04 & 6.1 \\
& \ion{Fe}{xii} & 195.12 & 6.1 \\
&               & 196.64  & 6.1 \\
& \ion{Fe}{xiii} & 196.54 & 6.2 \\
&                & 202.04 & 6.2 \\
&                & 203.82 & 6.2 \\
& \ion{Fe}{xiv} & 274.20 b &6.3 \\
&               & 264.79 & 6.3 \\
& \ion{Fe}{xv} & 284.16 & 6.3 \\

\noalign{\smallskip}
\multicolumn{4}{l}{Hot lines ($6.4\le \log\,T_{\rm max} <
  6.7$)} \\
\noalign{\smallskip}

& \ion{S}{xiii} & 256.68 b & 6.4 \\

& \ion{Fe}{xvi} & 262.98 & 6.4 \\

& \ion{Ni}{xvii} & 249.18 & 6.5 \\

&\ion{Fe}{xvii} & 254.87 & 6.6 \\
\noalign{\smallskip}
\multicolumn{4}{l}{Flare lines ($\log\,T_{\rm max} \ge 6.7$)}\\
\noalign{\smallskip}

& \ion{Ca}{xvii} & 192.82 b & 6.7 \\

& \ion{Fe}{xxiii} & 263.76 & 7.1 \\

& \ion{Fe}{xxiv} & 192.03 b & 7.2 \\
& \ion{Fe}{xxiv} & 255.11 b & 7.2 \\

\hline
\end{tabular}
\end{center}
\label{tbl.key-lines}
\end{table}

\section{Considerations for designing EIS observing studies}

Table~\ref{tbl.key-lines} summarises the emission lines recommended in
the previous sections.
Each ion in the Sun's atmosphere is formed over a characteristic
narrow temperature 
range, and the temperature of maximum abundance ($T_{\rm max}$) is
given in  Table~\ref{tbl.key-lines}. The recommended density
diagnostics described in the text are summarised in
Fig.~\ref{fig.dens}. 
The data rate
for EIS is sufficiently high  that 10--20 wavelength windows
can be chosen in most cases and thus many of the recommended lines can
be selected. For density diagnostics, we recommend the \ion{Fe}{xii}
\lam196.64/\lam195.12, and \ion{Fe}{xiii} \lam196.54/\lam202.04 and
\lam203.82/\lam202.04 ratios for all studies. For the other
diagnostics an assessment should be made based on the science to be
achieved (e.g., for transition region brightenings in active regions
the \ion{O}{v} and \ion{Mg}{vii} ratios should be selected).
The EIS planning software allows spectral
windows to have variable sizes which can be used to select windows
that span more than one emission line. This is useful for lines that
are close in wavelength (e.g., \ion{Fe}{viii} \lam194.66 and
\ion{Fe}{xii} \lam195.12) and for cases where a large window is
required to yield a good background measurement (e.g., \ion{Fe}{x}
\lam257.26).

Although the EIS emission lines typically extend over 12 wavelength
pixels 
we recommend setting the widths of wavelength windows to be 32 or 40
pixels (window sizes have to be multiples of 8 pixels) as the emission
lines have been seen to broaden significantly 
in flares and other dynamic phenomena. Broad windows also allow the
background level 
in the spectra to be more accurately measured.
If high cadence is required, then using 24 pixels to
reduce the data rate is acceptable, but high velocity events
may be missed. For \ion{Fe}{xii} \lam195.12 a window size of 48 pixels
(or larger) is recommended as the instrument is so sensitive at this
wavelength. 

The choice of exposure time for EIS studies depends on both the target
and science objective. If good signal is required in a wide range of
lines, then exposure times of 60--90~s for quiet Sun and 20--40~s for
active regions are recommended. If higher cadence is required, then
good signal in the strong lines (including the recommended
\ion{Fe}{xii} and \ion{Fe}{xiii} density diagnostics) can be obtained
in 20~s for quiet Sun and 5~s for active regions. Exposure times down
to 1~s will give a good signal in the \ion{Fe}{xii} \lam195.12 line in
active regions. These exposure times are for the 1$^{\prime\prime}$
slit -- they should be reduced by a factor 2 for the 2$^{\prime\prime}$
slit, and about a factor 2.5 for the slots.

\section{Summary}

The quality of the EIS spectra are outstanding and reveal a large
number of emission lines throughout the two wavelength bands that
offer exciting diagnostic opportunities, a number of which are
exploited in papers in this volume of PASJ. The present work has
summarised 
some of the key emission lines and density diagnostics based on 
evaluations of the initial EIS data-sets using the CHIANTI atomic
database. It is hoped that this will be a valuable reference for
scientists designing 
EIS studies, and help improve the science return from the instrument.

~

Hinode is a Japanese mission developed and launched by ISAS/JAXA, with
NAOJ as domestic partner and NASA and STFC (UK) as 
international partners. It is operated by these agencies in
co-operation with ESA and NSC (Norway). G.~Del Zanna and H.E.~Mason acknowledge
support from PPARC/STFC. The work of E.~Landi is supported by 
NASA. G.~Del Zanna thanks the
hospitality of DAMTP, University of Cambridge.

\end{document}